# High-Directional Wave Propagation in Periodic Gain/Loss Modulated Materials

N. Kumar[1*], M. Botey[2], R. Herrero[1], Yu. Loiko[3], and K. Staliunas[1,4]

[1]*Departament de Física i Enginyeria Nuclear, Universitat Politècnica de Catalunya, Colom 11, 08222 Terrassa, Spain*
[2]*Departament de Física i Enginyeria Nuclear, Universitat Politècnica de Catalunya, Urgell 187, 08036 Barcelona, Spain*
[3]*Departament de Física, Universitat Autònoma de Barcelona, E-08193, Bellaterra, Spain*
[4]*Institució Catalana de Recerca i Estudis Avançats (ICREA), Spain*

**Abstract**

Amplification/attenuation of light waves in artificial materials with a gain/loss modulation on the wavelength scale can be sensitive to the propagation direction. We give a numerical proof of the high anisotropy of the gain/loss in two dimensional periodic structures with square and rhombic lattice symmetry by solving the full set of Maxwell's equations using the finite difference time domain method. Anisotropy of amplification/attenuation leads to the narrowing of the angular spectrum of propagating radiation with wavevectors close to the edges of the first Brillouin Zone. The effect provides a novel and useful method to filter out high spatial harmonics from noisy beams.

*Keywords:* gain/loss modulated material, anisotropic gain, diffraction

## 1. Introduction

Light–matter interaction is one of the most addressed subjects in scientific research. In nature, the light–matter interaction occurs on the scale of single atom states [1] [2] [3] . Single atom response can be controlled by external fields and by varying the wavelength of propagating light. On the other hand, the response of a medium to light is a collective effect and can be controlled by the structure of the media, i.e. by the relative positions of different atoms within the material. Consequently, the propagation of light can be substantially modified in structured artificial materials: in metamaterials, where the magnetic and electric responses are modified on a nanometer scale [4] [5] , and also in Photonic Crystals (PhC), where the refractive index is modulated on the wavelength scale [6] [7] [8] .

During the last decade new beam propagation effects have been discovered and investigated in the field of PhCs. Precisely due to the periodic modulation of refractive index, the spatial dispersion curves −isofrequency lines of the Bloch modes, $\omega(k)$, or equivalently the curve families $k_\parallel$ ($k_\perp$) −are modified in comparison to those in homogeneous media, leading to different beam-propagation peculiarities: self-collimation [9] [13] , focalization and imaging behind the PhC structure [14] [1] [16] , spatial (angular) filtering [17] [18] , among others. We note that those spatial propagation effects are not restricted to electromagnetic waves but are also observed or predicted in other fields of wave mechanics, acoustics [19] [20] , and atom Bose condensates [24] .

As it is evident from the well-known Kramers–Kronig relations, a modulation of the refractive index is generally accompanied by a modulation of absorption or emission. Moreover, at resonant frequencies, the effects of absorption/emission can become relevant and have to be taken into account in addition to beam propagation effects arising from refractive index modulation. As a result, the new problem to be considered is light propagation within materials with a spatially modulated gain/loss. In the general case, one should consider simultaneously both gain/loss and index modulation. The present paper provides a detailed analysis of light propagation in Gain/Loss Modulated Materials (GLMMs) where purely the gain/loss is modulated.

Despite, in recent years, GLMMs have become more and more technologically relevant and accessible, they still do not attract the same interest as PhCs. The GLMMs with periodic modulation in one spatial dimension (1D) were considered in the context of distributed feedback lasers, where the effects of reflections of plane waves from 1D periodic structures is studied [25] [26] [27] . Other examples of 1D GLMMs are semiconductor lasers based on multiple quantum wells and semiconductor laser arrays [28] [29] ]. Recently, wave dynamics in 1D parity-time invariant potentials has been considered, which can also be regarded as systems with a particular combination of gain and refraction index modulation [30] [31] [32].

Recently, it has been proposed that GLMMs can provide interesting and technically useful beam propagation effects in two dimensional (2D) geometries [33] [34]. At the basis of these effects there is a high angular anisotropy of the gain/loss; i.e. a strong dependence of the gain/loss on the direction of propagation of electromagnetic waves. Apart from the above mentioned theoretical numerical approaches [33] [34], no other studies of light beam propagation, neither theoretical nor experimental have been reported in 2D GLMMs, to the best of our knowledge.

The first reported study on beam shaping effects was based on the paraxial approximation [33], therefore it cannot be directly applicable to GLMMs modulated on a wavelength scale, since possible reflections at large angles are excluded from the paraxial treatment. The basic predictions of Ref. [33] are: *i)* self-collimation (analogous to the effect observed in PhCs [9] ); *ii)* focalization by the structure −analogous to flat-lensing in PhCs [14] −; and *iii)* high directivity of the gain (having no direct correspondence in PhCs). The angular dependence of transparency (angular bandgaps), one of the most celebrated properties of PhCs, has also been proposed for spatial filtering of monochromatic beams [17] [18]. Such directivity in GLMMs has been also predicted by plane wave expansion (PWE) calculations in [34]. The isofrequency dispersion surfaces, ω($k$), not only show the distortion of the real part of the eigenfrequencies, $\omega_{re}(k)$, but also indicate the appearance of sharp peaks in the imaginary part of the eigenfrequencies, $\omega_{im}(k)$, which corresponds to an anisotropic gain in $k$-space. The present paper proves the predictions of Ref. [34] by numerical integration of the full set of Maxwell's equations using the Finite Difference Time Domain (FDTD) method. FDTD method is used for realistic simulations of light propagation through photonic structures, as a design tool, and contains no simplifications [35]. We perform numeric simulations in order to evaluate quantitatively the beam propagation effects for square and rhombic 2D lattices and, finally, to demonstrate the spatial (angular) filtering of noisy beams.

## 2. Model

We consider a 2D periodic structure made of lossy cylinders with either a square or rhombic (centred rectangular) symmetry embedded in an inactive background (no gain, no loss) as schematically shown in Fig. 1a and Fig. 1b. Therefore, the structure we investigate is absorbing on average. If gain is present in the structure waves amplitude can grow exponentially to infinity. In this case, nonlinear effects saturating the growth of light intensity should have to be taken into account in FDTD calculations; however, this is out of the scope of this study. For definiteness the radius, $R$, of the lossy cylinders is fixed to be $R = 0.2*a$, where $a$ denotes the lattice constant of the structure, and the absorption coefficient within the cylinders is fixed to be α = 50000 cm$^{-1}$, which is of the order of magnitude of absorption in semiconductor materials. This corresponds to complex-valued refractive index $n_{cyl}$ = 1+0.3979i. As already mentioned, in order to investigate pure effects caused by the loss-gain modulation the surrounding medium is considered to be air. The real part of the eigenfrequencies in Fig. 1c and Fig. 1d, correspond to frequencies of the Bloch modes and its imaginary part, shown in Fig. 1e and Fig. 1f, corresponds to the net gain or loss of each Bloch mode. In GLMMs, the dispersion curves of uncoupled modes do not push each other at the crossing points to form a band gap (which is the case for PhCs), but mutually pull and lock to a common frequency. The wave components at the edge of the first Brillouin Zone (BZ) lock to a common frequency to form a pattern of maximum gain (or minimum loss).

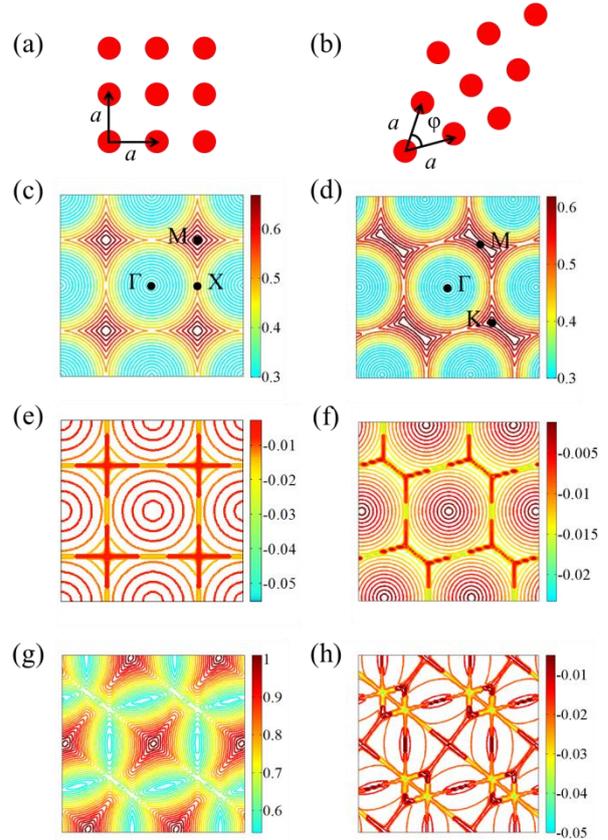

Fig. 1. (Color online) The structures considered have direct lattice vectors $\vec{a}_1 = a\vec{i}, \vec{a}_2 = a(\cos\varphi\, \vec{i} + \sin\varphi\, \vec{j})$ where φ = 90º for the square (a) and 60º<φ<90º for a rhombic geometry (b); as a particular case we consider φ = 75º. Figures (c)/(e) and (d)/(f) represent the isolines of the real/imaginary frequency, $\omega_{re}/\omega_{im}$, of the first band in a square and rhombic lattice, respectively and (g)/(h) represent the real and imaginary frequency of second band of rhombic lattice. The side intensity bar indicates real and imaginary parts of frequency in $a/\lambda$ units, where $a$ is the modulus of the lattice vectors.

The results throughout the paper contain simulations in purely lossy and linear materials, however the main conclusions hold for more general GLMMs. Modes with the lowest losses for the case of absorptive GLMMs correspond to modes with the highest gain for amplifying GLMMs. Therefore, modes surviving in lossy GLMMs, i.e. modes presenting the smallest net absorption have a similar spatial distribution with respect to modes with the highest amplification in a GLMM with a net gain.

## 3. Results

In most of the calculations we consider an input fundamental Gaussian light beam with transverse profile waist of 1.5 μm at the GLMM entrance surface. The electric field is considered to be linearly polarized parallel to the incidence plane of the structure; but similar results can be found for perpendicular polarization. In order to monitor the spatial dispersion and absorption properties of the GLMM under investigation, the near field transverse intensity distribution is calculated at a given propagation distance, $z \sim 2\, L_R$, where $L_R \approx 7$ μm stands for the Rayleigh length. We then obtain the far field profile by a spatial Fourier transformation in the transverse direction, and we determine the angular spectrum of absorption. We scan the frequency, and expect the most interesting effects to be observed at the edges of the 1st BZ, namely in the ΓM direction. As a reference for comparison, we also calculate the propagation of the same input beam through a homogenous material.

A typical result for a square structure is shown in Fig. 2. We note that the antinodes of this particular field correspond to the areas with minimum losses as indicated in Fig. 2e, which ensures the minimum overall loss of the Bloch mode. We obtain the complex-valued near field envelope (Fig. 2d) at a fixed distance from the entrance surface of the GLMM, indicated as a vertical line, D1, in Fig. 2a. Its Fourier transform, as shown Fig. 2e, exhibits a modulation due to the field harmonics arising from the periodicity of the GLMM in the transverse direction

### 3.1 Angular transmission profile

The angular transmission profile reconstructed from the field distributions behind the GLMM structure with a square lattice is presented in Fig. 3. We scan the frequencies around the edge of the 1st BZ, M point, and normalise the central part of the far field to the far field of a reference beam to obtain the angular transmission profile. The transmission profile, represented in Fig. 3a, can be compared with the imaginary part of the eigenfrequencies representation shown in Fig.1. The plot shows two symmetrically placed areas of minimum losses, above and below the edge of the 1st BZ, which denote the higher transmission angular regions. Such regions correspond to the resonance between two harmonics as shown in Figs. 3e,g. At the edge of the 1st BZ, full resonance is envisaged as all four relevant dispersion curves crossing at one point, as schematically shown in Fig. 3f. The side bands appearing in the spectral far field distributions below and above the M point, Fig. 3b and Fig. 3d, show the anisotropy of the gain which also causes the spectral narrowing close to the M point, see Fig. 3c.

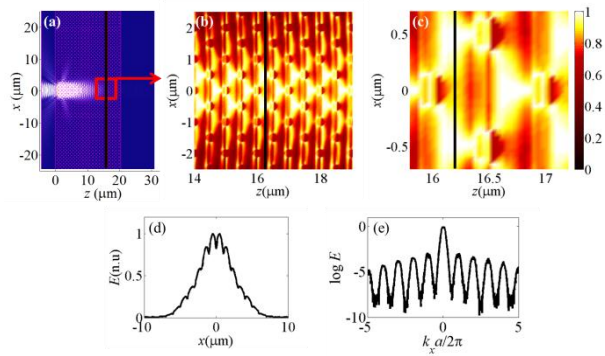

Fig. 2. (Color online) (a) Light beam propagation inside lossy GLMM with square lattice geometry, calculated by FDTD. The magnified view of the electric field shows the modulation of the propagated beam in the structure (b), and the shape of the Bloch mode (c). Vertical line D1 in figure (a) represents a plane at a distance z ∼ 2 $L_R$ used to reconstruct the angular transmission function. (d) Spatial distribution along the line D1 and (e) its Fourier transform.

The red lines in Fig. 3a represent the crossing of plane wave's dispersion curves in $(k_x, k_y)$ space. Whereas in PhCs such crossings indicate the appearance of band gaps, in GLMMs they coincide with enhanced gain areas. These crossing points can be easily located regarding the dispersion relations of the nearest harmonics. The dispersion relations are given by the circles $k_x^2 + k_y^2 = k^2$ for the central harmonic and $(k_x \pm q_x)^2 + (k_y - q_y)^2 = k^2$ for the two nearest harmonics, where vector $q = (q_x, q_y)$ is given by the lattice periodicity. The third nearest harmonic is centred at $q = (0, 2q_y)$. The crossing points of the central and the three nearest harmonics are obtained as a function of $k_x$:

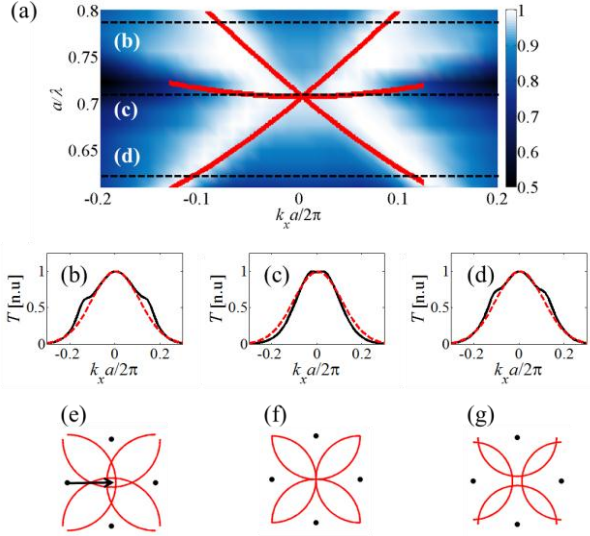
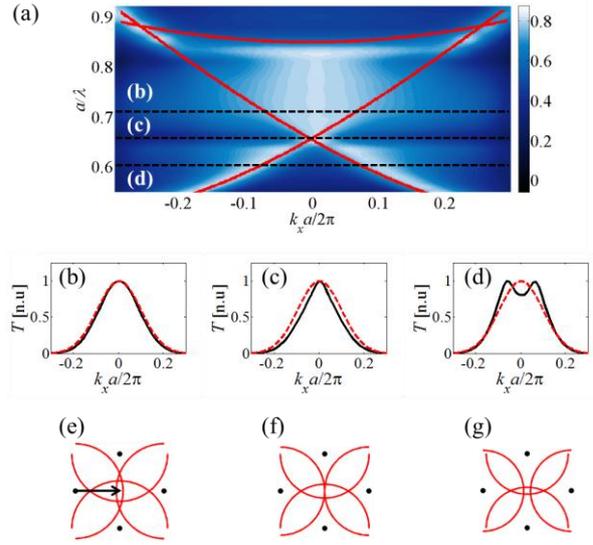

Fig. 3. (Color online) (a) Normalized angular transmission function for a square GLMM. The solid red curves correspond to the crossing points of the nearest harmonics given by eq. 1. The solid lines represent the far field distributions at D1 for frequencies: 0.63 (b), 0.71 (c) and 0.78 (d), in $a/l$ units, as a function of the normalised transverse component of the wavenumber. The dashed red lines represent the far field of the reference beam (propagating in free space). Figures (e), (f), (g) schematically represent the dispersion curves in the reciprocal lattice space, in absence of interaction for the frequencies considered above. The corner of the 1st BZ, (f), corresponds to resonance (crossing of the dispersion curves).

$$k = \sqrt{k_x^2 + \left(\frac{q_x^2 + q_y^2 \mp 2k_x q_x}{2q_y}\right)^2},$$
$$k = \sqrt{k_x^2 + q_y^2} \quad (1)$$

Similarly, the reconstruction of the transmission function has been performed for a GLMM with a rhombic geometry. Fig. 4 and Fig. 5 represent the analysis of the far field distributions for a rhombic lattice with $\varphi = 75°$, for light beams propagating along both, the long and the short diagonals of the direct lattice respectively, i.e. along the ΓK and ΓM crystallographic directions. It is important to note that the narrowing of the far field distribution is more pronounced for the rhombic geometry than for the square, as follows from the comparison of Fig. 4c and Fig. 5c with Fig. 3c.

### 3.2 Spatial filtering

The high directional gain in the vicinity of the edges of the BZ gives rise to spatial (or angular) filtering effects. For light beams with wave vectors close to the edge of the 1st BZ, the anisotropic loss profile around $k_\perp = 0$

Fig. 4. (Color online) Normalized transmission functions of a rhombic GLMM along ΓK direction. The solid red curves correspond to the crossing points of the nearest harmonics given by eq. 1. The K high symmetry point, at the edge of the 1st BZ, corresponds to the frequency 0.65. In figures (b), (c), (d) the far field distributions at D1 are plotted for the frequencies below (0.61), close (0.66) and above (0.70) the K point, as schematically represented in figures (e), (f) and (g)

results in the sole survival of the nearly on-axis modes, filtering out the high spatial harmonics and making the transverse profile smoother. For the analysis preformed in the previous section we infer that the most suitable structure for such purpose is the rhombic. In Fig. 6 the filtering process for a beam propagating along the long diagonal of a rhombic structure is demonstrated. An input beam with a noisy transverse profile is obtained by propagation of a Gaussian beam through a diffusive structure. The noisy components that correspond to the high spatial frequencies -or high transverse wavenumber- are gradually filtered out as shown in Fig. 6b in propagation through a 30 μm long GLMM. Therefore, behind the GLMM structure the light beam becomes of better spatial quality. The propagation of the filtered beam just behind the GLMM through a free space or homogeneous material enhance the filtering effect due to the rapid divergence of high spatial frequencies (sidebands in Fig. 6d) imposed by the spatial modulation of the material.

### 4. Conclusions

We have investigated propagation of light beams in two dimensional structures with periodic gain/loss modulation on a wavelength scale with square and rhombic symmetries. By means of FDTD method we have proved highly directive angular gain/loss profiles in the transmission, according to the predictions under paraxial approximation in Ref. [33] . Close to the edges of the first BZ, a narrowing of the spatial spectrum is observed. This effect can be related to the coupling and locking of spatial harmonic components propagating along ΓM direction. Besides, it is demonstrated that such strong anisotropy of losses can be used for spatial filtering purposes. We demonstrate the effect by filtering out the high angular components from the transverse profile of a noisy beam in relatively short propagation through the GLMM. Although angular narrowing of the gain profile is found for both square and rhombic geometries, the spatial filtering results to be more pronounced for a rhombic geometry and for light beams propagating along the long diagonal of the structure. The filtering effect of the GLMM structures demonstrated in this article could be used, for instance, to amplify and filter out, by means of optical methods, the information signals (wavelength/s) from the noisy background in optical communication systems either fiber based or free space.

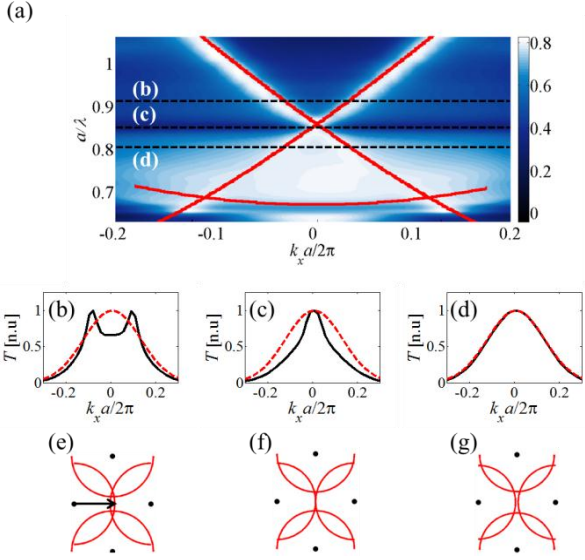

Fig. 5. (Color online) Normalized transmission functions of a rhombic GLMM along the ΓM direction. The solid red curves correspond to the crossing points of the nearest harmonics given by eq. 1, which intersect around the frequency 0.85. In figures (b), (c), (d) the far field distributions at D1 are plotted for frequencies below (0.82), close (0.85) and above (0.90) such point, as schematically represented in figures (e), (f) and (g).

## Acknowledgements

This work is financially supported by Spanish Ministerio de Educación y Ciencia and European FEDER through project FIS2008-06024-C03-02 and by COST action MP0702 from the European Community.

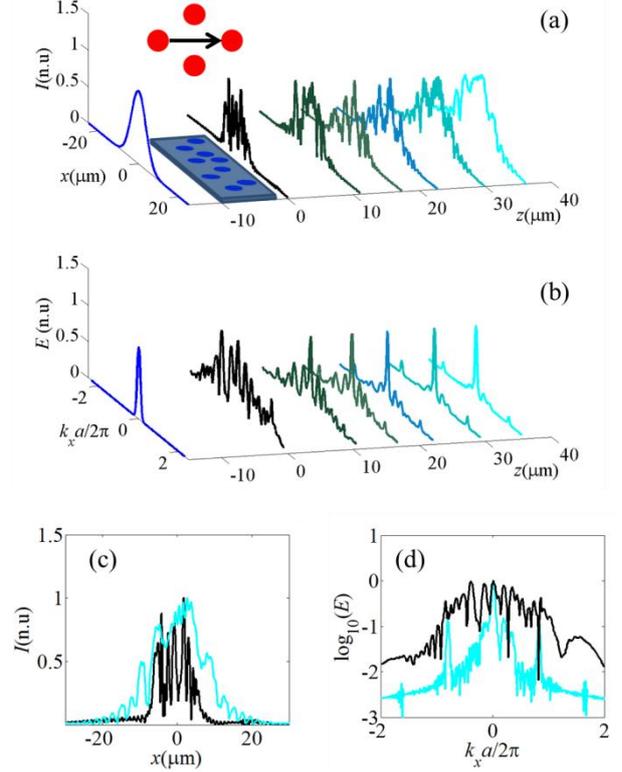

Fig. 6. (Color online) (a) Evolution of the beam profile trough the GLMM, propagating along the long diagonal. The noisy beam is generated by propagation of a Gaussian beam, with central frequency 0.85, through a disordered array of diffusers. (b) Spatial filtering effects −far field spatial Fourier spectra distribution− at different propagation distances. (c) Input noisy beam (dark solid curve) and output beam filtered by the structure (clear blue curve), and (d) their far field profiles, respectively.